\documentclass[11pt]{article}
\usepackage{hyperref}
\pdfoutput=1
\begin{document}
\title{The way to reduce electrical charge of \\ a droplet dispensed from a pipette tip}
\author{Dongwhi Choi, Horim Lee, Do Jin Im and Dong Sung Kim \\
\\\vspace{6pt} Mechanical Engineering, \\ POSTECH, KOREA}
\maketitle
\begin{abstract}
Recently, our group reported that an any aqueous droplet dispensed from a pipette tip has considerable amount of electrical charge. This natural electrical charge of a droplet could cause undesired, unfamiliar experimental results. Since the origin of the charge of a droplet is related to the pipette tip material, we modified the inside material of the pipette tip with poly(dimethylsiloxane)-graphene nanocomposites. The droplets dispensed from the modified pipette tip has lower electrical charge than that from the common pipette tip. We compare the experimental results with the droplets from common pipette tip and modified pipette tip. This fluid dynamics video includes the principle of the spontaneous electrification of the droplet and the results of the droplet in oil experiments.  
\end{abstract}
\section{Detailed explanation}
Our fluid dynamics video includes 4 segments:
\begin{enumerate}
\item The undesired, unfamiliar experimental results with droplet in oil system. 
\\  The used droplets were DI water droplet and the fibroblast cell suspension. The surrounding medium is silicone oil (50 cs). The charge of a DI water droplet dispensed from the pipette tip spontaneously has positive charge and the cell suspension in our experiment has negative charge. This cause a natural electrical repulsion among the DI water droplets and electrical attraction between cell suspension and the DI water droplet. Since this natural electrification phenomenon is not well-known, these experimental results could seem to be strange. 
\item The principle of the spontaneous electrification of the droplet dispensed from a pipette tip. 
\\  When the solid contact with aqueous solution, the natural electrification on the wall of the solid occurs. During dispensing the solution from a pipette tip, this generated charges are dispensed together with the solution.
\item The way we used to reduce the charge amount of a droplet. 
\\  We modified the pipette tip to reduce the charge amount. Since the origin of the charge is highly related to the inner surface material of the pipette tip, we coated the inner surface with polydimethylsiloxane-graphene nanocomposites through dip coating method.
\item Compared experimental results with the common, modified pipette tip. 
\\  We performed the experiment to confirm that the charge of the droplet dispensed from the modified pipette tip has lower amount than that from the common pipette tip. The merging of the DI water droplet dispensed from the modified pipette tip was easily achieved because of the lower electrical repulsion force among them. We compared the experimental results to show that the pipette tip material could have much effect on the droplet in oil experiment.
\end{enumerate}
\section{Reference}
Spontaneous electrical charging of a droplet dispensed by conventional pipetting, Scientific Reports, 3, 2037, (2013).
\end{document}